\documentstyle[12pt]{article} 
\begin{document}
\def\theequation{\arabic{section}.\arabic{equation}}
\newcommand{\be}{\begin{equation}} 
\newcommand{\ee}{\end{equation}}
\begin{titlepage} 
\title{A new solution for inflation} 
\author{Valerio Faraoni \\ \\
{\small \it Research Group in General Relativity (RggR)}\\
{\small \it Universit\'{e}
Libre de Bruxelles, Campus Plaine CP231}\\
{\small \it Boulevard du Triomphe, 1050 Bruxelles, Belgium}\\
{\small \it E--mail: vfaraoni@ulb.ac.be}\\
{\small and}\\
{\small \it INFN-Laboratori Nazionali di Frascati}\\
{\small \it P.O. Box 13, I-00044 Frascati, Italy}
} \date{} 
\maketitle
\thispagestyle{empty} 
\vspace*{1truecm} 
\begin{abstract} 
Many pedagogical  
introductions  of inflation are effective due to the simplicity of the relevant
equations. 
Here an analytic solution of the
cosmological equations is presented and used as an example 
to discuss fundamental aspects of the inflationary paradigm.
\end{abstract} 
\vspace*{1truecm} 
\begin{center} 
To appear in {\em Am. J. Phys.}
\end{center} 
\end{titlepage} 


\section{Introduction}

The standard big bang
cosmology$^{1-4}$ is a very
successful theory of
the universe with certain drawbacks. They include the horizon,
flatness and monopole problems$^{5,6}$   
or, in short, the ability to 
explain the observed universe only when  the
initial  conditions are extremely fine-tuned. Cosmic inflation solves the
problems of big bang cosmology 
and was introduced$^7$ at the beginning of the 80's. Today, almost 20
years
later, inflation is incorporated into modern cosmology and is a mature
topic
for inclusion in introductory courses in gravitation and cosmology. 

Inflation consists of a short period of accelerated superluminal 
expansion of the early universe, at the end of which the description of
the  
standard big bang model is applied. During the inflationary epoch, the
matter
content of
the
universe has an equation of state very close to that of the quantum
vacuum,
$P=-\rho$ (where $\rho $ and $P$ are, respectively, the energy density and
pressure of matter). Inflation also provides
a mechanism for the generation of  density perturbations through quantum
fluctuations of the scalar field which is supposed to drive the
cosmic expansion$^{4,8-12}$.

Although the original scenario of inflation and many others proposed to
date are based on specific
particle physics theories, the point of view of modern cosmology has
shifted: inflation is currently regarded as a {\em paradigm}, a general
idea that
can be implemented in a variety of ways to describe the early universe.
There are many inflationary
scenarios in the literature, but none is accepted as compelling and there
is no ``standard model'' of inflation. In this phenomenological point of
view it is possible to present cosmological inflation in the classroom
in
an easy way, without previous knowledge of advanced particle
physics. The task is made feasible by the simplicity of the
equations of inflation,  ordinary differential equations which are
solved in the slow rollover approximation, a simplification used in most
inflationary theories. 
Inflation is the subject of numerous textbooks and
introductory review papers$^{4,8,10-13}$ as
well as 
science popularization articles and books$^{14}$.

Important issues in understanding inflation are  
the slow rollover approximation, the equivalence
between a constant scalar field potential and the cosmological constant,
and the
correspondence between the equation of state and the scalar field
potential. This paper discusses these aspects of inflation in
detail, through the example of an exact analytic solution of the
dynamical equations. The latter is derived using elementary calculus and
is used to clarify the issues mentioned
above, with the added virtue of being useful
for introducing the cosmic no-hair theorems.

\section{The slow rollover approximation} 

\setcounter{equation}{0}

In this section we recall the basics about the cosmological constant,
inflation, and the slow rollover approximation used for simplifying and
solving the equations of inflation.

In general relativity a spatially
homogeneous and isotropic universe is
described by  a metric of the Friedmann-Lemaitre-Robertson-Walker
class$^{1-4}$. For simplicity,
we set our
discussion in such a universe  with flat spatial sections, described by
the line element
\be  \label{metric}
ds^2=-dt^2+a^2(t) \left(  dx^2 + d y^2+ d z^2 \right)  \; ,
\ee
in comoving coordinates $\left( t,x,y,z \right)$, where $a(t)$  is the
{\em scale factor} of the universe. For this metric, the Einstein
equations of general relativity reduce to$^{15-17}$ 
\be   \label{EF1}
\frac{\ddot{a}}{a}=-\, \frac{4\pi G}{3} \left( \rho_m +3P_m \right)
+\frac{\Lambda}{3}   \; ,
\ee

\be   \label{EF2}
H^2 \equiv \left( \frac{\dot{a}}{a} \right)^2= \frac{8\pi G}{3} \rho_m  
+\frac{\Lambda}{3}  \; ,
\ee
where an overdot denotes differentiation with respect to the comoving time
$t$ and $\rho_m$ and $P_m$ are, respectively, the energy density and
pressure 
of the material content of the universe, which is assumed to be a
perfect fluid. $H \equiv \dot{a}/a $ is the Hubble parameter, $\Lambda $
is the cosmological constant, $G$ is  Newton's constant
and units are used in which the speed of light in vacuum assumes the 
value unity. 

As is clear from the inspection of Eqs.~(\ref{EF1}) and (\ref{EF2}), it
is possible to describe the contribution of the cosmological constant as a
fluid with energy density and pressure given by
\be
\rho_{\Lambda}= \frac{\Lambda}{8\pi G} \; , \;\;\;\;\;\;\;\;\;\;
P_{\Lambda}=-\, \frac{\Lambda}{8\pi G} \; , 
\ee
respectively, and equation of state $P=-\rho$. Accordingly, one can
rewrite
Eqs. (\ref{EF1}) and (\ref{EF2}) as
\be   \label{EF11}
\frac{\ddot{a}}{a}=-\, \frac{4\pi G}{3} \left( \rho +3P \right) \; ,
\ee
\be   \label{EF22}
\left( \frac{\dot{a}}{a} \right)^2= \frac{8\pi G \rho }{3}  \; ,
\ee
where $ \rho \equiv \rho_m +\rho_{\Lambda}$ and $ P \equiv  P_m
+P_{\Lambda} $.  

{\em Inflation} is defined as an epoch in the history of the universe
during 
which the cosmic expansion is accelerated, $ \ddot{a} >0$.
Eq.~(\ref{EF11}) shows that acceleration is equivalent to a
negative pressure satisfying $P<-\rho/3$. It turns out that an
inflationary
period in which the universe expands by the factor e$^{70}$ solves the
fine-tuning problems of the standard big
bang cosmology$^{18}$.

In the original model and in most scenarios, inflation is
obtained
by assuming that at an early time of the order of $10^{-34}$~seconds the
energy density of the cosmological fluid was dominated by a scalar field 
called {\em inflaton}. Scalar fields are ever present in particle
physics, and it is natural that they played a role when the universe had 
the size of a subatomic particle. Inflation can then be seen as scalar
field
dynamics; the energy density and pressure of a scalar field $\phi$
minimally coupled to gravity are
given by$^{19,20}$ 
\be  \label{density}
\rho=\frac{( \dot{\phi})^2}{2} + V \; ,
\ee
\be  \label{pressure}
P=\frac{( \dot{\phi})^2}{2} - V \; ,
\ee
where $V( \phi ) $ and $ ( \dot{\phi})^2 /2 $ are, respectively, the 
potential and kinetic energy densities of the scalar field. The
scalar $\phi$ only depends on the time coordinate, due to the
assumption of
spatial homogeneity, and satisfies the well known Klein-Gordon
equation$^{2,4}$ 
\be  \label{KG} 
\ddot{\phi}+3H \dot{\phi} +\frac{dV}{d\phi}=0 \; .
\ee
By inserting Eqs.~(\ref{density}) and (\ref{pressure})
into Eqs.~(\ref{EF11}) and (\ref{EF22}) one obtains 
\be  \label{EI1}
\frac{\ddot{a}}{a}=\frac{8\pi G}{3} \left[ V( \phi ) -\dot{\phi}^2 \right]
\; ,
\ee
\be  \label{EI2}
H^2=\frac{8\pi G}{3} \left[ V( \phi ) + \frac{\dot{\phi}^2}{2} \right] \;
,
\ee
which, together with Eq.~(\ref{KG}), constitute the {\em equations of 
inflation}. 
Note that only two of the three equations (\ref{KG}), (\ref{EI1}) and
(\ref{EI2}) are independent; when $\dot{\phi}\neq 0$, Eq.~(\ref{KG})
follows from the energy conservation equation$^{21}$ 
\be \label{conservation}
\dot{\rho}+3H \left( \rho + P \right) =0 \; .
\ee
Different inflationary scenarios correspond to different
choices of the form of the scalar field potential $V( \phi )$,
which are usually motivated by particle physics arguments. Certain
scenarios are set in theories of gravity alternative to general
relativity$^{22}$, and will not be considered here.

From the didactical point of view, it is interesting that the dynamics of
the inflaton field can be viewed as
the motion of a ball with unit mass, position $\phi$ and speed
$\dot{\phi}$ rolling on a hill which has a profile
given by$^4$ the shape of $V( \phi )$. 

The equations of
inflation are usually
solved in the {\em slow rollover approximation}, which
assumes that the inflaton's speed $\dot{\phi}$ is small, that $V( \phi )
\gg \dot{\phi}^2/2$, and that the inflaton's motion (described by
Eq.~(\ref{KG})) is friction-dominated. Then, the acceleration term
$\ddot{\phi}$ can be
neglected in comparison with the force term $V' \equiv dV/d\phi$ and with
the friction term $3H \dot{\phi}$ in Eq.~(\ref{KG}). A necessary condition
for this to occur is that the inequalities
\be  \label{sr1}
\left| \frac{V'}{V} \right| \ll 4 \sqrt{\pi G} \; ,
\ee
\be  \label{sr2}
\left| \frac{V''}{V} \right| \ll  8 \pi G \; ,
\ee
hold$^{23}$. In other words, in the slow rollover 
approximation
the {\em slow roll
parameters}
\be  \label{epsilon}
\epsilon \equiv \frac{1}{16\pi G} \left( \frac{V'}{V} \right)^2 \; ,
\ee
\be  \label{eta}
\eta \equiv  \frac{1}{8\pi G} \frac{V''}{V}  \; ,
\ee
are small, $\epsilon, |\eta | \ll 1$. $\epsilon $ and $\eta$ are related,
respectively, to the slope and the curvature of the scalar field potential  
$V( \phi)$;  the smallness of $\epsilon$ and $| \eta |$ means that the
potential $V(
\phi )$ is relatively flat during slow roll inflation.
The equations of inflation
reduce to the two first order equations
\be
H^2=\frac{8\pi G}{3} \, V \; ,
\ee
\be
3H \dot{\phi}=- \frac{dV}{d\phi}  \; .
\ee
It would be misleading to put excessive
emphasis on
the requirement that $\epsilon$ and $\eta$ be small and on the
existence of an almost flat section of the potential $V( \phi )$ on which
the scalar field
$\phi$ can slowly roll  (i.e. with small speed $\dot{\phi}$). 
Indeed, Eqs.~(\ref{sr1}) and (\ref{sr2}) are necessary but not sufficient
conditions for neglecting the acceleration term in Eq.~(\ref{KG}); a
solution of the field equations may well satisfy Eqs.~(\ref{sr1})
and (\ref{sr2}), but still have a large speed $\dot{\phi}$ and therefore
not be in slow-rollover; this is the case, at early times, for the exact
solution presented in Sec.~4.

\section{Equivalence between a constant potential and the cosmological
constant}

\setcounter{equation}{0}

In this section we discuss the de Sitter solution viewed as the prototype
of
inflation; the similarity between slow roll inflation and the de
Sitter-like 
expansion of the universe is emphasized.

The de Sitter solution of the Einstein equations corresponds to
exponential cosmic expansion,
\be  \label{dS}
a(t)=a_0 \, \mbox{e}^{Ht}\; , 
\ee
\be
H=\left( \frac{\Lambda}{3} \right)^{1/2} \; ,
\ee
and is obtained by  setting $\rho_m=P_m=0
$ in Eqs.~(\ref{EF1}) and (\ref{EF2}) and by allowing only the
cosmological constant in the right hand side of the field equations.
Historically, the de Sitter
solution has been known since the beginnings of general relativistic
cosmology, long before the idea of inflation. As seen in the
previous section, the cosmological constant is equivalent to a fluid with
the equation of state of quantum vacuum
\be  \label{vacuum}
P=-\rho \; ,
\ee
and therefore cosmological constant and vacuum energy are synonyms in
modern cosmology$^4$. The equation of state
(\ref{vacuum}) uniquely leads to the solution (\ref{dS}). The idea that
quantum vacuum should
be considered as a form of matter and hence as a source of gravity in the
Einstein equations arose in the Russian school$^{24}$. 

While the original de Sitter solution (\ref{dS}) is associated to a
geometrical cosmological constant, matter can generate an
effective cosmological constant without the need of introducing a
geometrical $\Lambda$. In fact, the vacuum energy of
a scalar field mimics a geometric cosmological constant and the de Sitter
expansion of the universe (\ref{dS}) can be obtained as the result
of scalar field dynamics; the latter are rather trivial. A {\em
constant} scalar field
\be   \label{constantphi}
\phi =\phi_0 \; ,
\ee
with the potential $ V=V_0 = $constant 
solves the equations of inflation (\ref{EI1}), (\ref{EI2}), and 
(\ref{KG}) and
\be
a=a_0 \exp \left( \sqrt{ \frac{8\pi G V_0}{3}} \, t \right)
\ee
for this solution. The idea that a constant, or almost constant, scalar
field
appearing in grand 
unified theories of particle physics plays the role of  a vacuum state
contributed to the development of the idea of inflation$^{24}$.

The de Sitter solution was the prototype of inflation$^{4,8,11,12.14}$,
but
its importance
for inflation is not merely historical. In fact, most of the inflationary
scenarios proposed thus far share the common feature of being solved in
the slow rollover approximation (\ref{sr1}) and (\ref{sr2}), which implies
that
the expansion of the universe is quasi-exponential during the slow
rollover
phase. In fact, when the scalar field is in slow rollover, the dominance
of
the potential over the kinetic energy density, $(
\dot{\phi})^2 /2 \ll V $, implies that $\rho \approx V $ and $P \approx -V
$, and the equation of state is {\em approximately} (\ref{vacuum}), which
is equivalent to say that the scale factor is approximated by (\ref{dS}).
More precisely, during slow roll inflation,
the scale factor is given by
\be
a(t)=a_0 \exp \left( \int H(t) dt \right) \simeq a_0 \exp \left(
H_0t +\frac{ H_1}{2}\, t^2 \right) \; ,
\ee
where the constant term $H_0$ in the expansion of $H(t)$ is dominant.

In the rest of this paper we set the geometric cosmological constant to
zero and we only consider the effective cosmological constant due to the
scalar field. 

\section{A new analytic solution}

\setcounter{equation}{0}

A new analytic solution can be derived in the classroom using only
elementary calculus, and is presented in this section. It differs from
the de Sitter solution, but it is derived from Eqs. (\ref{EF1}), 
(\ref{EF2}), (\ref{density}), and (\ref{pressure}) under the same
assumption of constant scalar field potential that led to
the de Sitter solution (\ref{dS}).

One begins by noting that during slow roll inflation the nearly flat
section of the scalar field potential plays the role of an effective 
cosmological constant. One then sets out to determine all the
solutions of the {\em exact} equations of inflation
(\ref{KG})-(\ref{EI2}) corresponding to a
constant scalar field potential
\be
V=V_0=\mbox{const.} 
\ee
The Klein-Gordon equation (\ref{KG}) reduces to 
\be   \label{reducedKG}
\ddot{\phi}+3H\dot{\phi}=0 \; , 
\ee
a trivial solution of which is given by $\dot{\phi}=0$, and  corresponds
to the de Sitter solution discussed in the
previous section. Another solution is possible when $\dot{\phi}\neq 0$; in
this case Eq.~(\ref{reducedKG}) can be divided by $\dot{\phi}$ and
immediately integrated to yield
\be  \label{phidot}
\dot{\phi}=\pm \, \frac{C}{a^3} \; ,
\ee
where $C$ is a positive integration constant. By substituting
Eq.~(\ref{phidot}) into Eq.~(\ref{density}) and the
resulting expression of the energy density into
Eq.~(\ref{EF2}) with $\Lambda=0$, one obtains
\be   \label{deltadelta}
\left( \frac{\dot{a}}{a} \right)^2=\frac{8\pi G}{3} \left(
\frac{C^2}{2a^6}+V_0 \right)
\; .
\ee
Upon use of the auxiliary variable $y\equiv \ln a$, Eq.~(\ref{deltadelta})
can be reduced to a quadrature,
\be
\int \frac{dy}{\sqrt{1+\alpha \exp{(  -6y)}}}=\pm \sqrt{\frac{8\pi G
V_0}{3}}  \left( t-t_0 \right) \; ,
\ee
where $\alpha = C^2 / ( 2V_0) $ and $t_0$ is an integration constant. By
using 
\be
\int \frac{dz}{z\sqrt{1+z}}=\ln \left( \frac{\sqrt{z+1}-1}{\sqrt{z+1}+1}
\right) \; ,
\ee
where $ z=\alpha $e$^{-6y}=\alpha a^{-6}$, one easily obtains
\be   \label{threedots}
\frac{\sqrt{a^6+\alpha}-a^3}{\sqrt{a^6+\alpha}+a^3}=\exp \left( 
\mp 12\sqrt{\frac{2\pi G V_0}{3}} \, t \right) \; ,
\ee
where the boundary condition $a( t=0 ) =0$ has been imposed. 
The function
\be
f(x)=\frac{\left( x^6+\alpha \right)^{1/2}-x^3}{\left( x^6+\alpha
\right)^{1/2} +x^3}
\ee
is monotonically decreasing for $x \geq 0$ and, since $f(0)=1$, it is
$0<f<1 $ for any $x> 0$. The right hand side of Eq.~(\ref{threedots})
is
always greater than unity when the positive sign is adopted, hence
Eq.~(\ref{threedots}) has no solutions in this case. By contrast, adopting
the negative sign in Eq.~(\ref{threedots}), one obtains after
straightforward calculations
\be  \label{anew}
a(t)=a_1 \sinh^{1/3} \left( \sqrt{3\Lambda}\,  t \right) \; , 
\ee
where $a_1=\left( 4\pi G C^2 /\Lambda \right)^{1/6} $ and $ \Lambda =8\pi
G V_0$. Eq.~(\ref{phidot}) yields the corresponding scalar field
\be   \label{phinew}
\phi (t) =\phi_0 \ln \left[ \tanh \left( \frac{\sqrt{3\Lambda}\, t}{2} 
\right) \right]  +\phi_1 \; ,
\ee
where $\phi_0=\pm \left( 12\pi G \right)^{-1/2} $ and $\phi_1$ is an
integration constant with the dimensions of
a mass.
It is straightforward to check that the solution (\ref{anew}) and
(\ref{phinew}) satisfies Eqs.~(\ref{KG}) and (\ref{EI2}); to check that
Eq.~(\ref{EI1}) is also satisfied, it is sufficient to note that, given
Eqs.~(\ref{KG}) and (\ref{EI2}), Eq.~(\ref{EI1}) follows from the
conservation equation (\ref{conservation}).

The solution given by Eqs. (\ref{anew}) and (\ref{phinew}) has a big bang
singularity at $t=0$, with the asymptotic behavior 
\be
a(t) \approx t^{1/3} \; ,\;\;\;\;\;\; \phi \approx \phi_0 \ln \left(
\frac{\sqrt{3\Lambda}\, t}{2} \right)  +\phi_1 
\ee
as $t \rightarrow 0$. At $t=0$ the scale factor vanishes while
$\phi$, $\rho$ and $P$ diverge. The initial speed $\dot{\phi}$ also
diverges and the solution is definitely not in slow rollover, despite the
fact
that the potential is flat (in fact, constant). However, a slow rollover
regime is approached as the universe evolves: at late times
$t\rightarrow +\infty$ the solution (\ref{anew}) and 
(\ref{phinew}) is asymptotic to the de Sitter solution (\ref{dS}). This is
in agreement with the cosmic no-hair theorems, as explained later.

\section{Equation of state and scalar field potential}

\setcounter{equation}{0}

We will now return to the exact solution (\ref{anew}) and (\ref{phinew})
and interpret them in terms of the equation of state relating pressure and
energy density.

The effective equation of state of the universe corresponding 
to Eqs.~(\ref{anew}) and (\ref{phinew}) is given by
\be
\frac{P}{\rho}= 1-2\tanh^2 \left(  \sqrt{3\Lambda}\, t \right) \; ;
\ee
it changes with time and interpolates between the equation of
state
$P=\rho$ at early
times and the vacuum equation of state $P=-\rho$  at late times.  This
feature of the solution cautions against a possible
misunderstanding, i.e. the belief that fixing
the scalar
field potential $V( \phi )$ is equivalent to prescribing the equation of
state. This belief would be false: in fact, by fixing the scalar field
potential to be constant, $V=V_0$, one does not uniquely determine the
evolution of the universe: the solutions (\ref{anew}), (\ref{phinew}) and
(\ref{dS}), (\ref{constantphi}) 
correspond to very different physical situations and equations of state.
In order to completely specify the microphysics, it is not sufficient to
prescribe the scalar field potential, but one must provide complete
information on the 
state of the scalar field (i.e. also the field's ``speed'' $\dot{\phi}$ in
our
example).

For a {\em general} potential $V$, the effective equation of state of the
universe dominated by a scalar field is given by
\be
\frac{P}{\rho}=\frac{ \dot{\phi}^2 -2V}{ \dot{\phi}^2 +2V} \equiv w(x) \;
,
\ee
where $x \equiv \dot{\phi}^2 /2V $ is the ratio between the kinetic and
the
potential energy densities of the scalar $\phi$. Under the usual
assumption $V \geq 0$ (which guarantees that the energy density $\rho$
is non-negative when $\dot{\phi}=0$), one has that, for $x \geq 0$,
the function $w(x)=\left(x^2-1\right) \left( x^2+1 \right)^{-1} $
increases monotonically from its minimum $w_{min}=-1$
attained at $x=0$ to
the horizontal asymptote $+1$ as $x \rightarrow + \infty$. The slow
rollover 
regime corresponds to the region $|x | \ll 1$ and to $w$ near its minimum,
where the kinetic energy density of $\phi$ is negligible in comparison to
its potential energy density. As the kinetic energy density
$\dot{\phi}^2/2 $
increases, the equation of state progressively deviates from $P=-\rho$ and
the pressure becomes less and less negative; the system gradually moves
away from the slow rollover regime. At the equipartition between
the kinetic and the potential energy densities ($x=1$), one
has the ``dust'' equation of state $P=0$. The pressure becomes positive as 
$x$ increases and, when the kinetic energy density
completely dominates the potential energy density ($x \gg 1$), one
finally reaches 
the equation of state $P=\rho$. The solution (\ref{anew}) and 
(\ref{phinew}) for $V=$const. spans the entire possible range for the
equation of state during the evolution of the universe, starting from
$x=+ \infty$ at early times and asymptotically evolving towards $x=0$ at
late times.

\section{Discussion and conclusions}

\setcounter{equation}{0}

We now comment on the exact solution described by Eqs.~(\ref{anew}) and 
(\ref{phinew}) and compare it to the de Sitter solution. In doing so, we
will sharpen our understanding of the slow rollover approximation and
introduce an interesting cosmic no-hair theorem.

Both solutions (\ref{anew}), (\ref{phinew}) and (\ref{dS}),
(\ref{constantphi}) of the Einstein-Friedmann equations correspond to a
constant scalar field potential $V=V_0$ and {\em exactly} satisfy Eqs. 
(\ref{sr1}) and
(\ref{sr2}), which are necessary, but not sufficient, conditions for
slow roll inflation. While
the de Sitter solution (\ref{dS}) corresponds to perfect slow roll
($\dot{\phi}=0$) and to exact exponential expansion of the universe, the
solution
(\ref{anew}), (\ref{phinew}) exhibits significant differences. It has a
big bang singularity at $t=0$, while the de Sitter universe with infinite
age has existed forever and the timescale over which the latter changes,
$\sqrt{3/\Lambda}$, is constant (see Ref.~25 for a pedagogical discussion
of the self-similarity properties of the de Sitter solution).
By contrast, the solution given by Eqs.~(\ref{anew}) and  
(\ref{phinew}) has a timescale given by the Hubble time 
\be
H^{-1}=\sqrt{\frac{3}{\Lambda}}\, \tanh \left( \sqrt{3\Lambda}\, t \right)
\;,
\ee
which is time-dependent and becomes nearly constant (with value
$\sqrt{3/\Lambda}$) only for $t \rightarrow + \infty$. Near the big bang
singularity
the universe obeys the equation of state $P=\rho$. In this
region, not only the solution fails to be in slow roll over the flat
potential, but it is not even inflationary; even though the slow roll
parameters $\epsilon$ and $\eta$ {\em exactly} vanish, the speed
$\dot{\phi}$ of the scalar field is large. This clearly shows
that the slow rollover approximation must be formulated by means of
conditions on the {\em solutions} of the field equations, not only as a
set of  conditions on the scalar field {\em potential} $V( \phi )$. 
The relative flatness of the potential is not all there is to
the slow rollover approximation.

Finally, the solution described by Eqs.~(\ref{anew}) and (\ref{phinew})
constitutes a useful example
to introduce the cosmic no-hair theorems$^{26}$. The
latter 
state that, when a positive cosmological constant (or vacuum energy)
$\Lambda$ is present, the de Sitter solution (\ref{dS}),
(\ref{constantphi}) behaves as
an attractor for the other
solutions$^{26}$. The example solution (\ref{anew})
and 
(\ref{phinew}) clearly illustrates this behavior and is particularly
useful to introduce  the cosmic no-hair theorems (the proof of which
requires more sophisticated mathematics than the ones used in
this paper$^{26}$). In spite of the fact
that the solution (\ref{anew}) and (\ref{phinew}) begins at early times 
very differently from
a de Sitter solution, it converges exponentially fast to the latter 
as time progresses. In fact, the ratio of the (suitably normalized) scale
factors
(\ref{anew}) and (\ref{dS}) is given by $\left[ 1-\exp
\left( -2\sqrt{3\Lambda} \, t \right)  \right]^{1/3} $, which tends to
unity at large times $t$, while the scalar
field (\ref{phinew}) asymptotically converges to the constant
$\phi_1$. This happens because the  de Sitter solution (\ref{dS}) is an
attractor point that captures the orbits of the solutions in the phase
space, including the orbit of the exact solution (\ref{anew}) and
(\ref{phinew}). Indeed, the cosmic no-hair theorems are more general,
stating that the de Sitter space is approached even starting from an
anisotropic spacetime$^{26}$.

As a conclusion, cosmological inflation is described by simple ordinary
differential equations and its basic features can be discussed in the
classroom without the need of complicated mathematical tools. The
phenomenological approach to inflation adopted in recent years by the
community of cosmologists  supports the introduction of the
inflationary paradigm without the need of a lengthy premise about advanced
particle physics to justify it. Indeed, the basic features of
inflation can be grasped without the knowledge of high energy theories.

The
increasing number of pedagogical introductions to
inflation seems to reflect this point of view. The exact solution shown
in this paper should help the circulation of simple but significant
ideas of modern cosmology.

\clearpage

\section*{Acknowledgments}

The author thanks L. Niwa for reviewing the manuscript and an anonymous
referee for valuable suggestions.   
This work was supported by the EEC grants PSS*~0992 and SEAC-1999-00039
and by OLAM, Fondation
pour la Recherche Fondamentale, Brussels. 

\vskip1.5truecm

\noindent $^1$  L.D. Landau and E.M. Lifschitz, {\em The
Classical Theory of Fields} (Pergamon, Oxford, 1989), pp.~358-368.\\
$^2$  R.M. Wald, {\em General Relativity} (Chicago
University Press, Chicago, 1984), pp.~91-117.\\
$^3$ R. D'Inverno, {\em Introducing Einstein's Relativity}
(Clarendon, Oxford, 1992), chap.~22-23.\\
$^4$ E.W. Kolb and M.S. Turner, {\em The Early Universe} 
(Addison--Wesley, Reading, MA, 1990), pp.~261-317.\\
$^5$ R.M. Wald, {\em op. cit.}, pp.~107-108.\\
$^6$ E.W. Kolb and M.S. Turner, {\em op. cit.}, pp.~261-269.\\
$^7$ A.H. Guth, {\em The Inflationary Universe: A Possible Solution to the
Horizon and Flatness Problems}, Physical Review D {\bf 23}, 347-356.\\
$^8$ A.D. Linde, {\em Particle Physics and Inflationary
Cosmology} (Hardwood Academic, Chur, Switzerland, 1990).\\
$^9$ A.R. Liddle and D.H. Lyth, {\em The Cold Dark
Matter Density Perturbation}, Physics Reports {\bf 231}, 1--105 (1993).\\
$^{10}$ M.S. Turner, in {\em Recent Directions in
Particle Theory~-~From Superstrings and Black Holes to the Standard
Model}, Proceedings of the Theoretical Advanced Study Institute in
Elementary
Particle Physics, Boulder, Colorado 1992, J. Harvey and J. Polchinski eds.
(World Scientific, Singapore, 1993).\\
$^{11}$ S.K. Blau and A.H. Guth, {\em Inflationary Cosmology}, in {\em
300 Years of Gravitation}, 
S.W. Hawking and W. Israel eds. (Cambridge University Press, Cambridge,
1987), pp.~424-603.\\
$^{12}$ A.R. Liddle, {\em An Introduction to Cosmological Inflation}, in
Proceedings of the ICTP Summer School in High Energy Physics, Trieste,
Italy 1998, in press (also available as preprint astro-ph/9901124 on the
Fermilab preprint server).\\
$^{13}$ J.E. Lidsey, A.R. Liddle, E.J. Copeland, T.
Barreiro and M. Abney, {\em Reconstructing the Inflaton Potential--an
Overview}, Reviews of Modern Physics {\bf 69}, 373-410 (1997).\\
$^{14}$ A.H. Guth, {\em The Inflationary Universe} (Addison-Wesley,
Reading, MA 1997).\\ 
$^{15}$ R.M. Wald, {\em op. cit.}, p.~97.\\
$^{16}$ R. D'Inverno, {\em op. cit.}, p.~322.\\
$^{17}$ E.W. Kolb and M.S. Turner, {\em op. cit.}, pp.~49-50.\\
$^{18}$ E.W. Kolb and M.S. Turner, {\em op. cit.}, pp.~281-283.\\
$^{19}$ R.M. Wald, {\em op. cit.}, p.~70.\\
$^{20}$ E.W. Kolb and M.S. Turner, {\em op. cit.}, p.~277.\\
$^{21}$ R.M. Wald, {\em op. cit.}, p.~100.\\
$^{22}$ A.R. Liddle and D.H. Lyth, {\em op. cit.}, pp.~52-54.\\
$^{23}$ A.R. Liddle and D.H. Lyth, {\em op. cit.}, pp.~42-43.\\
$^{24}$ A.D. Linde, {\em op. cit.}, pp.~35-36.\\
$^{25}$ R. D'Inverno, {\em op. cit.}, pp.~337-338.\\
$^{26}$ E.W. Kolb and M.S. Turner, {\em op. cit.}, pp.~303-309.

\end{document}